# An Investigation on Social Network Recommender Systems and Collaborative Filtering Techniques

Maryam Nayebzadeh[1], Akbar Moazzam[2], Amir Mohammad Saba[1], Hadi Abdolrahimpour[3], Elham Shahab[1]
Department of Computer Engineering, Azad Islamic University, Yazd, Iran[1]
{mnayebzadeh, amsaba, ma.shahab}@iauyazd.ac.ir
Department of Computer Science, Isfahan University, Isfahan, Iran[2]
Department of Bio Mechanic, Azad Islamic University, Yazd, Iran[3]

*Abstract*—Nowadays, with the remarkable expansion of the information through the internet, users prefer to receive the exact information that they need through some suggestions from their friends or profiles to save their time and money. Recommend systems based on different algorithms as one of the basic ways to reach this goal through the internet have been proposed but each of them has their own advantages and disadvantages. In this study, we have selected and implemented two approaches which are Collaborative Filtering (CF) and Social Network Recommendations System (SNRS). Based on some limitations to finding a dataset which covers friendship, rating and item categories we generated it for 10 categories, 10 items, and 100 users and compared two approaches. We used Mean Absolute Error (MAE) and accuracy to compare the result of two mentioned approaches and found that the SNRS method as it is claimed to be improved version of CF works more efficiency.

*Index Terms*— Collaborative Filtering, Social Network Recommendations System.

## I. Introduction

Finding relevant information has been a challenge for a long time. Based on this requirement, recommendation systems are becoming one of the most popular technique to encounter with this problem. Almost, all recommendation system try to make predictions about the preference of each user based on the preferences of a set of similar users through content-based filtering, collaborative filtering or combination of both. Collaborative filtering is the prevalent recommendation system which has been used to identify users that can be characterized as a similar according to the logged history of prior behaviors. Generally, a collaborative filtering algorithm use a collection of user profiles to identify interesting information, items, products, services or etc. for these users. A particular user achieves a recommendation based on the user profiles of other similar users. User profiles are made by asking users to rate items, items, products, services or etc. There are several applications for collaborative filtering in different domains such as trust and security [10,19,23,31], web services recommender systems [14, 32], and context-aware recommenders [13, 20]. In this paper, we have compared a traditional collaborative filtering with the improved version of probabilistic approach (which is known as Social Network-based Recommender System).

## II. Related Work

Nowadays, recommender systems are becoming one of the approaches that help users to make decision in regards of what products to buy, which news to read and what movie to watch. These systems try to maintain the loyalty of users and increase sale from producer's point of view and on the other side save user's time and money through proposing the most matched recommendations to users.

Generally, three categories for recommender systems can be enumerated which are collaborative filtering, content-based filtering and the hybrid version which is the combination of two mentioned categories. Social recommender systems which are known as improved version of collaborative filtering are based on social networks. Social networks are made of a finite group of users and their relationships (figure1) that they establish among them through the social links so one key insight is that social-based recommender systems should account for a number of dimensions within a user's social network, including social relationship strength, expertise, and user similarity. There are several techniques to extracting information from social graph network such as information retrieval [9] and knowledge extraction [4] that can be applicable in collaborative filtering. Additionally, combining summarization techniques [17] and data mining approaches [15] with collaborative filtering model is getting more attention these days to receive a better accuracy. Analyzing the data generated by users within social networks has several practical applications that can be used to develop recommendation systems [2].

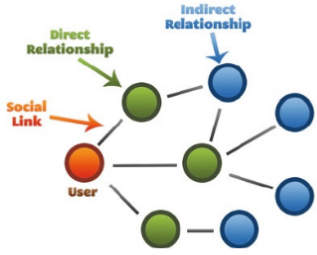

Figure 1: the pattern of social relationship in social networks [3].

### III. COLLABORATIVE FILTERING

Collaborative Filtering is a technique used by many recommendation systems. We can define Collaborative filtering as a method of making automatic predictions (filtering) about the interests of a user by collecting preferences or taste information from many users (collaborating)[4]. Collaborative Filtering can be of three types

*1) Memory Based Filtering:* In this approach we use ratings of the users, directly to the rating of the active user (user for we are predicting the rating to recommend). In most of the cases we do not use all the users in prediction. Instead of that we use a set of nearest neighbors and that's why this algorithm is known as "User-Based Nearest Neighbor Algorithm".

*2) Model Based Filtering:* Here system does not use the user ratings directly. Instead of that they form a model using those rating values and use that to predict the rating for the active user. This model can be formed using Bayesian Networks, Clustering, Association Rule Mining or Latent Semantic. All of them try to get a model that reduces the data to compute for each recommendation.

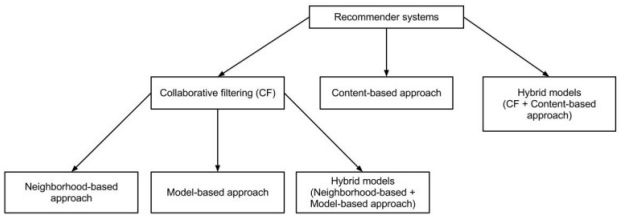

Figure 2: Recommendation system models[4].

*3) Hybrid Filtering:* Some applications use the previous two models together. This approach helps to overcome the short comings of both the algorithms. It improves the prediction performance. Importantly, it overcomes the CF problems such as sparsity and loss of information. However, they have increased complexity and are expensive to implement [5].

### IV. PROBABILISTIC APPROACH

Probabilistic Approach is a relatively new technique. As described in the paper, to predict that a user will like/dislike a given item, there are 3 factors to be considered.

*1) User Preference,* which is the probability of the current user to give a given rating to any similar item as of target item.

*2) Item Acceptance,* which the probability of the current item to get a given rating by all the users under consideration.

*3) Friend Inference,* which is the probability of the friends of current user to give a given rating to any similar item as of target item.

### V. REQUIREMENTS OF DATASET

Dataset needed for testing both of this algorithms needs lots more information than just social network and individual reviews. We tried many datasets namely yelp.com, facbook, movielens etc. But those datasets were not rich enough to test this algorithm. In the paper also they have made their own dataset with the help of information from yelp.com. The main requirements for the dataset were:

a. It must show social relationship bonds between different individuals.

b. It must show individual ratings given by individuals to different items.

c. It must show the categories to which the different items belong to. In other words, it must show the different attributes each of the different item have.

Since we cannot find a dataset good enough for this algorithm, we generated our own dataset.

### VI. DATASET AND IT'S GENERATION

To generate the dataset, we used Microsoft Excel. The basic heuristic that we used is that a user will give more or less similar rating to a given item as his/her friends gave to that same item. We needed 3 tables/sheets to store the needed data.

*1) Relationship Table*

In this, we stored the relationship bonds between different individuals of a social network. We took 100 users, so this table has 100 x 100 entries. Lets assume R(x,y) represents the bond between 2 users x & y, then

$R(x,y) = \{0,1,2,3,4,5\}$  x,y $\epsilon$ Users

$R(x,y) = R(y,x)$     ,x≠y, x,y $\epsilon$ Users

$R(x,y) =$ undefined   , x=y, x,y $\epsilon$ Users

The value of R(x,y) lies between 0 to 5, where 0 indicates extreme dislike and 5 extreme indicates best friends x & y. We assigned random ratings from 0 to 5 to random pairs of friends in this sheet. Some pairs were not.

rated at all, since it is not necessary that every individual knows every other individual in a social network.

*2) Rating Table*

In this, we stored the rating given by different individuals to different items. Part of this table was used as training dataset along with two of other tables and part of this table was used as testing dataset. In this we took 100 users & 10 items, so this table has 100 x 10 entries. Lets assume  K(x,y)  represents the rating given by user x to item y, then

$K(x,y) = \{0,1,2,3,4,5\}$

$K(x,y) \neq K(y,x)$    $x \in$ Users, $y \in$ Items

The value of $K(x,y)$ lies between 0 to 5, where 0 indicates extreme dislike and 5 extreme extreme likeness of a user x for a given item y. We generates this table manually by first taking some random individuals and putting random ratings for them against random items. Lets assume we put a random rating R from a user U to an item I. Now for all other users apart from this random ones, we calculates their ratings as average of what his/her friends gave to that item.

For example, User U2 has not been assigned any rating for given item I5, so we will find all friends of U2, who have given ratings to I5. Lets assume there are 3 friends of U2 namely U9, U17, U31 who have given rating R9, R17, R31 to item I5. So to calculate the rating that U2 will give to I5, we take the weighted average of the relationship between U2 & his friends & the ratings they gave to I5.

$K(U2, I5) = \sum R(U, X_i) \times K(X_i, I5) / \sum R(U, X_i)$

$i = \{9, 17, 31\}$.

*3) Attribute/Category Table*

In this, we stored the attributes each item has. In other words, we stored the categories to which an item belongs to. Also an item can have more than 1 attributes or it can belong to more than 1 categories. Lets assume $B(x,y)$ represents whether an item x belongs to category y or not, then

$B(x,y) \neq B(y,x)$    $x \in$ Items, $y \in$ Categories

$B(x,y) = \{0,1\}$

The value of $B(x,y)$ can be 0 or 1, where 0 indicates that item x doesn't belong to category y and 1 indicates that item x belongs to category y. In this we took 10 items and 10 categories, so this table has 10 x 10 entries. This entries were randomly generates.

VII. IMPLEMENTATION OF COLLABORATIVE FILTERING APPROACH

We have used the Memory based approach to implement our Model.

$$pred(u,i) = \sum_{n \subset neighbours} userSim(u,n) \cdot r_{ni}$$

This is the basic formula for the Memory based model. 'u' is the active user, 'I' is the item for which we are predicting the rating. 'userSim' is the function of calculating the similarity between two users(neighbors), $r_{ni}$ is the raring of user 'n' for item 'i'. Here 'n' is a neighbor of the active user.

This formula has many problems like:
- The prediction will be miss-scaled, if the similarities of the neighbors do not sum up to one. To calculate a weighted average of ratings, the sum of the coefficients that multiply the ratings should be 1. Otherwise:
  - If the sum of the similarities of the neighbours is higher than one, the predicted value will be higher than the real weighted average.
  - If the sum of the similarities of the neighbours is lower than one, the predicted value will be lower than the real weighted average.
- Different users use different rating scales in their minds. Is necessary to normalize the different scales [6].

Now we have come up to the formula:

$$pred(u,i) = \bar{r}_u + \frac{\sum_{n \subset neighbours (u)} userSim(u,n) \cdot (r_{ni} - \bar{r}_n)}{\sum_{n \subset neighbours (u)} userSim(u,n)}$$

- u, is the user.
- n, the neighbour to compare.
- i, is a item.
- $r_{ui}$ is the rating of the neighbour n to the item i.
- $r_{ni}$ is the rating of the neighbour n to the item i.
- $\bar{r}_n$ is the average of rating of the neighbour.
- $\bar{r}_u$ is the average of rating of the user.

To resolve the problem of the wrong scale due to the sum of the *userSim* coefficients, the result is divided by the sum of all coefficients to normalize the scale. Finally, to get the correct estimation the average value of the ratings of the user is added, because it was subtracted during the calculation.

$$userSim(u,n) = \frac{\sum_{i \subset CR_{u,n}} (r_{ui} - \bar{r}_u)(r_{ni} - \bar{r}_n)}{\sqrt{\sum_{i \subset CR_{u,n}} (r_{ui} - \bar{r}_u)^2} \sqrt{\sum_{i \subset CR_{u,n}} (r_{ni} - \bar{r}_n)^2}}$$

$Cr_{u,n}$ denotes the set of co-rated items between Active user u and his/her neighbors n.

This is the Pearson Correlation and it has been used to measure the correlation between the ratings of the Active user and its neighbor.

VIII. DISADVANTAGES OF COLLABORATIVE APPROACH

The basic challenge in collaborative filtering is named Cold-Start which in new items, products, services or etc. that have not been yet rated by the users. In this scenario many people have to rate the items, products, services or etc. before the system will be effective. Data sparsity is the another problem which can occur when there are many items, products, services or etc. to rate and user normally only rates a few items, products, services or etc. which it leads to impede finding users similar to the target one.

## IX. IMPLEMENTATION OF PROBABILISTIC APPROACH / SOCIAL NETWORK-BASED RECOMMENDER SYSTEM

The improved version of collaborative filtering tries to propose a model that takes user preference, item acceptance and friend inference into consideration (figure 1). We briefly explain each criterion and how they can be involved in final decision.

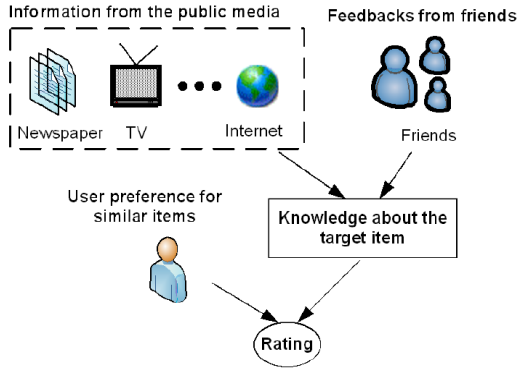

Figure 3: The three factors that influence a customer's buying decision[1].

1) User preference: user preference probability which is based on naïve Bayes assumption will be achieved through equation (1) where Pr(A'1, A'2, ..., A'n) and Pr(RU = k) are normalization constant and the prior probability that U gives a ranking k. User Preference is actually the probability that a user U will give a rating K to any item I provided that item has attributes A.

$$\Pr(R_U = k \mid A' = a'_I) = \frac{\Pr(R_U = k) \times \Pr(A'_1, A'_2, ...., A'_n \mid R_U = k)}{\Pr(A'_1, A'_2, ...., A'_n)}$$

2) Item acceptance: this probability which is similar to estimation in user preference, captures the general acceptance of item I from point of view users like user U. Item acceptance is the probability that a given item I will get rating K from users, provided that item has attributes A.

$$\Pr(R_I = k \mid A = a_U) = \frac{\Pr(R_I = k) \times \Pr(A_1, A_2, ...., A_m \mid R_I = k)}{\Pr(A_1, A_2, ...., A_m)}$$

where Pr(RI = k) is the prior probability that the target item I receives a rating value k, and Pr(Aj | RI = k) is the conditional probability that user attribute Aj of a reviewer has a value of aj given item I receives a rating k from this reviewer[1]. These two probabilities can be learned by counting the review ratings on the target item I in a manner similar to what we did in learning user preferences[1].

3) Friend inference: in this step, SNRS learns how the particular user U and each his/her friend V are related based on the items they have both rated previously. This probability is achieved through estimating user similarities or coefficients, either based on user profiles or user rating. Friend Inference for a user U for an Item I, is the probability that friends of that user will give a rating K to that item I.

4) To get the final probability, that a user U will give rating K to a given item I, provided that the item has attributes A, we multiply all three of them and divide by a normalizing factor. Z can be calculated from training dataset.

$P_f = (P_u \times P_i \times P_{fi}) / Z$

5) To calculate the predicted rating Kf, that a user U will give to an item I, provided that item I has attributes A, is given by taking the weighted mean of all rating from 1 to 5 with their corresponding probabilities.

$K_f = \sum P_{fi} * k_i / \sum P_{fi}$

i = 1,2,3,4,5.

Table 1: Actual dataset

|  | I1 | I2 | I3 | I4 | I5 |
|---|---|---|---|---|---|
| **U51** | 1 | 1 | 2 | 1 | 2 |
| **U52** | 1 | 4 | 2 | 0 | 4 |
| **U53** | 1 | 2 | 1 | 1 | 1 |
| **U54** | 1 | 2 | 1 | 2 | 3 |
| **U55** | 2 | 2 | 1 | 0 | 2 |
| **U56** | 2 | 3 | 1 | 1 | 1 |
| **U57** | 2 | 2 | 1 | 1 | 3 |
| **U58** | 3 | 1 | 1 | 1 | 1 |
| **U59** | 0 | 3 | 1 | 4 | 4 |
| **U60** | 2 | 2 | 4 | 1 | 1 |

Table 2: Recommended dataset from collaborative filtering approach.

|  | I1 | I2 | I3 | I4 | I5 |
|---|---|---|---|---|---|
| **U51** | 1 | 2 | 1 | 1 | 2 |
| **U52** | 2 | 2 | 2 | 2 | 2 |
| **U53** | 2 | 2 | 2 | 2 | 1 |
| **U54** | 1 | 1 | 1 | 1 | 1 |
| **U55** | 1 | 2 | 1 | 1 | 1 |
| **U56** | 1 | 2 | 2 | 2 | 2 |
| **U57** | 2 | 2 | 2 | 2 | 2 |
| **U58** | 1 | 2 | 1 | 1 | 1 |
| **U59** | 1 | 1 | 1 | 1 | 1 |
| **U60** | 1 | 2 | 1 | 2 | 1 |

Table 3: Original Data

|  | I6 | I7 | I8 | I9 | I10 |
|---|---|---|---|---|---|
| **U51** | 3 | 1 | 0 | 1 | 2 |
| **U52** | 2 | 4 | 2 | 1 | 1 |
| **U53** | 2 | 3 | 1 | 0 | 2 |
| **U54** | 2 | 2 | 0 | 1 | 1 |
| **U55** | 1 | 3 | 2 | 1 | 3 |
| **U56** | 2 | 2 | 0 | 1 | 2 |
| **U57** | 2 | 1 | 0 | 4 | 2 |
| **U58** | 2 | 2 | 1 | 0 | 1 |
| **U59** | 2 | 1 | 1 | 2 | 1 |
| **U60** | 2 | 2 | 0 | 0 | 1 |

Table 4: Recommended dataset from probabilistic approach.

|  | I6 | I7 | I8 | I9 | I10 |
|---|---|---|---|---|---|
| **U51** | 2 | 2 | 2 | 2 | 2 |
| **U52** | 1 | 1 | 1 | 1 | 1 |
| **U53** | 2 | 2 | 2 | 2 | 2 |
| **U54** | 2 | 2 | 2 | 2 | 2 |
| **U55** | 2 | 2 | 2 | 2 | 2 |
| **U56** | 2 | 2 | 2 | 2 | 2 |
| **U57** | 2 | 2 | 2 | 2 | 2 |
| **U58** | 2 | 2 | 2 | 2 | 2 |
| **U59** | 2 | 2 | 2 | 2 | 2 |
| **U60** | 2 | 2 | 2 | 2 | 2 |

## X. COMPARISON

We have used two measures to compare both approaches.

*1) MAE (Mean Absolute Error)*

Mean absolute is the mean of absolute error of the number of observations. In this project we compared 50 x 5 values. So total number of observations are 250.

$$MAE = \sum | R_i - A_i | / N$$

where $R_i \in$ Recommended Values

$A_i \in$ Actual Values

N = number of observations

*2) Accuracy*

Accuracy is the percentage of total number of observations in which the recommended made was exactly the same as the actual value.

$$Accuracy = (C / N) \times 100 \%$$

Where C is total number of observations in which the recommended made was exactly the same as the actual value.

N is total number of observations.

## XI. RESULTS

|  | MAE | Accuracy |
|---|---|---|
| CF | 0.876 | 35.2% |
| SNRS | 0.930 | 33.6% |

## XII. CONCLUSION

In this paper we have implemented two algorithms which are collaborative filtering and improved version of probability approach. As it was expected, the improved version which is known as Social Network Recommender System has better efficiency in comparison with traditional collaborative filtering because in this approach user's own preference, item's general acceptance and influences from friends have been taken into consideration. The main issue about the results is that the difference between two results are not that much noticeable which is somehow related to the limitations about finding actual data set based on two algorithms input requirements.

REFERENCES

[1] He, Jianming, and Wesley W. Chu. "A social network-based recommender system (SNRS)." In Data Mining for Social Network Data, pp. 47-74. Springer US, 2010.

[2] Hanneman, Robert A., and Mark Riddle. "Introduction to social network methods." (2005).

[3] Al Falahi, Kanna, Nikolaos Mavridis, and Yacine Atif. "Social networks and recommender systems: a world of current and future synergies." In Computational Social Networks, pp. 445-465. Springer London, 2012.

[4] E. D. Trippe, J. B. Aguilar, Y. H. Yan, M. V. Nural, J. A. Brady, M. Assefi, S. Safaei, M. Allahyari, S. Pouriyeh, M. R. Galinski, J. C. Kissinger, and J. B. Gutierrez. 2017. A Vision for Health Informatics: Introducing the SKED Framework.An Extensible Architecture for Scientific Knowledge Extraction from Data. ArXiv e-prints (2017). arXiv:1706.07992

[5] Breese, John S., David Heckerman, and Carl Kadie. "Empirical analysis of predictive algorithms for collaborative filtering." In Proceedings of the Fourteenth conference on Uncertainty in artificial intelligence, pp. 43-52. Morgan Kaufmann Publishers Inc., 1998.

[6] Billsus, Daniel, and Michael J. Pazzani. "Learning Collaborative Information Filters." In Icml, vol. 98, pp. 46-54. 1998.

[7] Koren, Yehuda, and Robert Bell. "Advances in collaborative filtering." In Recommender systems handbook, pp. 77-118. Springer US, 2015.

[8] Zhang, Fuzheng, Nicholas Jing Yuan, Defu Lian, Xing Xie, and Wei-Ying Ma. "Collaborative knowledge base embedding for recommender systems." In Proceedings of the 22nd ACM SIGKDD international conference on knowledge discovery and data mining, pp. 353-362. ACM, 2016.

[9] Zhang, Qian-Ming, An Zeng, and Ming-Sheng Shang. "Extracting the information backbone in online system." PloS one 8, no. 5 (2013): e62624.

[10] Doroodchi, Mahmood, Azadeh Iranmehr, and Seyed Amin Pouriyeh. "An investigation on integrating XML-based security into Web services." In *GCC Conference & Exhibition, 2009 5th IEEE*, pp. 1-5. IEEE, 2009.

[11] Heckerman, David E., John S. Breese, Eric Horvitz, and David Maxwell Chickering. "Collaborative filtering utilizing a belief network." U.S. Patent 5,704,017, issued December 30, 1997.

[12] Yang, Xiwang, Yang Guo, Yong Liu, and Harald Steck. "A survey of collaborative filtering based social recommender systems." Computer Communications 41 (2014): 1-10.

[13] Duan, Lian, W. Nick Street, and E. Xu. "Healthcare information systems: data mining methods in the creation of a clinical recommender system." Enterprise Information Systems 5, no. 2 (2011): 169-181.

[14] Blake, M. Brian, and Michael F. Nowlan. "A web service recommender system using enhanced syntactical matching." In Web Services, 2007. ICWS 2007. IEEE International Conference on, pp. 575-582. IEEE, 2007.

[15] M. Allahyari, S. Pouriyeh, M. Assefi, S. Safaei, E. D. Trippe, J. B. Gutierrez, and K. Kochut. 2017. A Brief Survey of Text Mining: Classification, Clustering and Extraction Techniques. ArXiv e-prints (2017). arXiv: 1707.02919

[16] Arazy, Ofer, Nanda Kumar, and Bracha Shapira. "Improving social recommender systems." IT professional 11, no. 4 (2009).

[17] M. Allahyari, S. Pouriyeh, M. Assefi, S. Safaei, E. D. Trippe, J. B. Gutierrez, and K. Kochut. 2017. Text Summarization Techniques: A Brief Survey. ArXiv e-prints (2017). arXiv:1707.02268

[18] Dooms, Simon, Toon De Pessemier, and Luc Martens. "Movietweetings: a movie rating dataset collected from twitter." In Workshop on Crowdsourcing and human computation for recommender systems, CrowdRec at RecSys, vol. 2013, p. 43. 2013.

[19] Pouriyeh, Seyed Amin, Mahmood Doroodchi, and M. R. Rezaeinejad. "Secure Mobile Approaches Using Web Services." In *SWWS*, pp. 75-78. 2010.

[20] Allahyari, Mehdi, and Krys Kochut. "Semantic Context-Aware Recommendation via Topic Models Leveraging Linked Open Data." In *International Conference on Web Information Systems Engineering*, pp. 263-277. Springer International Publishing, 2016.

[21] Su, Xiaoyuan, and Taghi M. Khoshgoftaar. "A survey of collaborative filtering techniques." *Advances in artificial intelligence* 2009 (2009): 4.

[22] Shi, Yue, Martha Larson, and Alan Hanjalic. "Collaborative filtering beyond the user-item matrix: A survey of the state of the art and future challenges." *ACM Computing Surveys (CSUR)* 47, no. 1 (2014): 3.

[23] Canny, John. "Collaborative filtering with privacy." In *Security and Privacy, 2002. Proceedings. 2002 IEEE Symposium on*, pp. 45-57. IEEE, 2002.

[24] Polat, Huseyin, and Wenliang Du. "Privacy-preserving collaborative filtering using randomized perturbation techniques." In *Data Mining, 2003. ICDM 2003. Third IEEE International Conference on*, pp. 625-628. IEEE, 2003.

[25] López-Nores, Martín, Yolanda Blanco-Fernández, José J. Pazos-Arias, and Alberto Gil-Solla. "Property-based collaborative filtering for health-aware recommender systems." *Expert Systems with Applications* 39, no. 8 (2012): 7451-7457.

[26] Sarwar, Badrul, George Karypis, Joseph Konstan, and John Riedl. "Item-based collaborative filtering recommendation algorithms." In *Proceedings of the 10th international conference on World Wide Web*, pp. 285-295. ACM, 2001.

[27] Huang, Zan, Hsinchun Chen, and Daniel Zeng. "Applying associative retrieval techniques to alleviate the sparsity problem in collaborative filtering." *ACM Transactions on Information Systems (TOIS)* 22, no. 1 (2004): 116-142.

[28] Shahab, Elham. "A Short Survey of Biomedical Relation Extraction Techniques." *arXiv preprint arXiv:1707.05850* (2017).

[29] Herlocker, Jonathan L., Joseph A. Konstan, Loren G. Terveen, and John T. Riedl. "Evaluating collaborative filtering recommender systems." *ACM Transactions on Information Systems (TOIS)* 22, no. 1 (2004): 5-53.

[30] E. Shahab. 2017. A Short Survey of Biomedical Relation Extraction Techniques. ArXiv e-prints (2017). arXiv: 1707.05850

[31] Massa, Paolo, and Bobby Bhattacharjee. "Using trust in recommender systems: an experimental analysis." In International Conference on Trust Management, pp. 221-235. Springer, Berlin, Heidelberg, 2004.

[32] A. M. Saba, E. Shahab, H. Abdolrahimpour, M. Hakimi, and A. Moazzam. 2017. A Comparative Analysis of XML Documents, XML Enabled Databases and Native XML Databases. ArXiv e-prints (2017). arXiv: 1707.08259

.